\documentclass[twocolumn,superscriptaddress,showkeys]{revtex4}
\usepackage{amssymb,epsf,epsfig,amsmath,color}
\usepackage{multirow}
\usepackage{accents}

\newcommand{\lsim}{
\mathrel{\hbox{\rlap{\hbox{\lower4pt\hbox{$\sim$}}}\hbox{$<$}}}}
\newcommand{\gsim}{
\mathrel{\hbox{\rlap{\hbox{\lower4pt\hbox{$\sim$}}}\hbox{$>$}}}}

\usepackage{xspace}

\newcommand{\Bbar}{\kern 0.18em\overline{\kern -0.18em B}{}\xspace}

\newcommand{\Kbar}{\kern 0.18em\overline{\kern -0.18em K}{}\xspace}

\def\theabstract{Exploring correlations between the CP asymmetries of $B^0_d\to\pi^0K_{\rm S}$ following 
from an isospin relation, we uncover new tensions with the Standard Model in data for neutral $B\to\pi K$ decays. 
Should this intriguing picture originate from New Physics, a modified electroweak penguin sector provides a key 
scenario. It includes models with extra $Z'$ bosons, which offer attractive ways to resolve anomalies in 
$B\to K^{(*)}\ell^+\ell^-$ measurements. We present a new strategy to reveal the underlying physics, apply it to 
current $B\to\pi K$ data, and discuss the excellent prospects for Belle II. 
}
\begin{document}
\begin{titlepage}
\vspace*{-0.1truecm}
\begin{flushright}
Nikhef-2017-045\\
SI-HEP-2017-28\\
QFET-2017-23
\end{flushright}

\vspace{1.3truecm}

\begin{center}
\boldmath
{\Large{\bf Towards New Frontiers with $B\to\pi K$ Decays}}
\unboldmath
\end{center}

\vspace{1.2truecm}

\begin{center}
{\bf Robert Fleischer,\,${}^{a,b}$  Ruben Jaarsma\,${}^{a}$ and  K. Keri Vos\,${}^{c}$}

\vspace{0.5truecm}

${}^a${\sl Nikhef, Science Park 105, NL-1098 XG Amsterdam, Netherlands}

${}^b${\sl  Department of Physics and Astronomy, Vrije Universiteit Amsterdam,\\
NL-1081 HV Amsterdam, Netherlands}

${}^c${\sl Theoretische Physik 1, Naturwissenschaftlich-Technische Fakult\"at, \\
Universit\"at Siegen, D-57068 Siegen, Germany}

\end{center}

\vspace*{1.7cm}

\begin{center}
\large{\bf Abstract}\\

\vspace*{0.6truecm}

\begin{tabular}{p{13.5truecm}}
\theabstract
\end{tabular}

\end{center}

\vspace*{5.1truecm}

\vfill

\noindent
December 2017

\end{titlepage}

\newpage
\thispagestyle{empty}
\mbox{}

\newpage
\thispagestyle{empty}
\mbox{}

\rule{0cm}{23cm}

\newpage
\thispagestyle{empty}
\mbox{}

\setcounter{page}{0}

\preprint{Nikhef-2012-nnn}

\date{\today}

\title{\boldmath Towards New Frontiers with $B\to\pi K$ Decays
\unboldmath}

\author{Robert Fleischer}
\affiliation{Nikhef, Science Park 105, NL-1098 XG Amsterdam, Netherlands}
\affiliation{Department of Physics and Astronomy, Vrije Universiteit Amsterdam,
NL-1081 HV Amsterdam, Netherlands}

\author{Ruben Jaarsma}
\affiliation{Nikhef, Science Park 105, NL-1098 XG Amsterdam, Netherlands}

\author{K. Keri Vos}
\affiliation{Theoretische Physik 1, Naturwissenschaftlich-Technische Fakult\"at, Universit\"at Siegen, 
D-57068 Siegen, Germany}

	\date{\today}
	\vspace{3em}

\begin{abstract}
\vspace{0.2cm}\noindent
\theabstract 
\end{abstract}


\maketitle

\thispagestyle{empty}
%
%
%
\setcounter{page}{1}
Decays of $B$ mesons offer a powerful tool for testing the Standard Model (SM). Particularly interesting probes 
are given by $B\to\pi K$ decays, which have entered the experimental stage about two decades ago and have 
also received a lot of attention from the theory community, with puzzling patterns in previous data 
\cite{BFRS-1,BFRS-2,FRS}. These decays are dominated by QCD penguin topologies, while tree diagrams are 
strongly suppressed by the element $|V_{ub}|$ of the Cabibbo--Kobayashi--Maskawa (CKM) matrix. 

In the case of the $B^0_d\to \pi^- K^+$ and $B^+\to \pi^+ K^0$ channels, the electroweak (EW) penguin topologies 
are color-suppressed and hence expected to play a minor role. On the other hand, $B^0_d \to \pi^0 K^0$ and 
$B^+\to \pi^0 K^+$ have also color-allowed EW penguin topologies, which contribute at the same level as the 
color-allowed tree amplitudes \cite{RF-96}. 

To characterize the EW penguin effects, we introduce
\begin{equation}
q e^{i\phi} e^{i\omega} \equiv - \left(\frac{\hat P_{EW} + \hat P_{EW}^{\rm C}}{\hat{T} +\hat{C}} \right),
\end{equation}
where $\phi$ and $\omega$ are CP-violating and CP-conserving phases, 
and $\hat P_{EW}$ ($\hat{T}$) and $\hat P_{EW}^{\rm C}$ ($\hat{C}$) denote color-allowed and color-suppressed 
EW penguin (tree) amplitudes, respectively. This parameter is very favourable from the theoretical point of view
and is of key interest for our analysis. It can be calculated by applying the $SU(3)$ flavor symmetry to the relevant 
hadronic matrix elements, yielding the following SM expression \cite{BF-98,NR} (see also \cite{RF-95}):
\begin{equation}\label{q-SM}
\hspace*{-0.1truecm}q e^{i\phi} e^{i\omega} = 
\frac{-3}{2\lambda^2 R_b}\left(\frac{C_9 + C_{10}}{C_1 + C_2}\right) R_q= (0.68 \pm 0.05) R_q,
\end{equation}
where $\lambda\equiv |V_{us}| = 0.22$, $R_b= 0.39\pm 0.03$ is a side of the unitarity triangle (UT), the $C_k$ are 
short-distance coefficients \cite{BBNS}, and $R_q$ may differ from 1 through $SU(3)$-breaking corrections. For the
analysis of current data, we use $R_q=1.0\pm0.3$ \cite{FJPZ}. We observe that the strong phase $\omega$ vanishes 
in the $SU(3)$ limit. The smallness of $\omega$ is a model-independent feature, as discussed in detail in 
Ref.~\cite{BBNS}, finding values at the few-degree level. 

The EW penguin sector opens an exciting avenue for New Physics (NP) to enter $B\to\pi K$ decays 
(see, for instance, Refs.~\cite{BFRS-1,BFRS-2,FRS,BBNS,FJPZ,HSV,BGV}).
Prominent scenarios are given by models with extra $Z'$ bosons 
\cite{BEJLLW-2,BEJLLW-1}. In view of anomalies in data for rare $B\to K^{(*)}\ell^+\ell^-$ 
decays, such SM extensions are receiving a lot of attention (see, e.g., Ref.~\cite{ANSS}). 
Should this kind of NP actually be realized in Nature, it would also affect the $B\to \pi K$ system, thereby 
putting these modes again into the spotlight \cite{RF-FPCP,BDLRR,FT}. 

Interestingly, we find that the current $B\to\pi K$ data result in a puzzling situation. Thanks in particular to sharper 
measurements of $\gamma$, a previous tension of CP violation in $B^0_d\to\pi^0K_{\rm S}$ with respect to the 
SM \cite{FJPZ} has become more pronounced. Moreover, we show that another probe offered by $B^0_d\to\pi^-K^+$
does not agree with its SM prediction. This intriguing situation could actually be resolved through NP in the 
EW penguin sector. 

Concerning this $B\to\pi K$ puzzle, $B^0_d\to\pi^0K^0$ plays a key role as it is the only channel with a 
``mixing-induced" CP asymmetry $S^{\pi^0K_{\rm S}}_{\rm CP}$. This observable arises from interference 
between $B^0_d$--$\bar B^0_d$ mixing and decay \cite{RF-95}:
\begin{displaymath}
\frac{\Gamma(\bar{B}_d^0(t) \rightarrow \pi^0K_{\rm S}) - 
\Gamma(B_d^0(t) \rightarrow \pi^0K_{\rm S})}{\Gamma(\bar{B}_d^0(t) \rightarrow \pi^0K_{\rm S}) + 
\Gamma(B_d^0(t) \rightarrow \pi^0K_{\rm S})} 
\end{displaymath}
\vspace*{-0.3truecm}
\begin{equation}\label{ACP}
= A^{\pi^0K_{\rm S}}_{\rm CP} \cos(\Delta M_dt) + S^{\pi^0K_{\rm S}}_{\rm CP}\sin(\Delta M_dt).
\end{equation}
Here the time dependence comes from $B^0_d$--$\bar B^0_d$ oscillations, with a frequency given by the mass 
difference $\Delta M_d$ between the $B_d$ mass eigenstates.

\begin{table}[t]
	\centering
	\begin{tabular}{l | r r  r }
		Mode & ${\mathcal Br} [10^{-6}]$& $A_\text{CP}^f$ & $S_\text{CP}^f$ 	 \\
		\hline\hline 	{	\vspace{-0.3cm}}\\
		${B}^0_d\to \pi^- K^+$ & $19.6\pm 0.5$  & $ -0.082 \pm 0.006$ & $-$\\
		${B}^0_d\to \pi^0 {K}^0$ & $9.9\pm 0.5$  &$ 0.00 \pm 0.13$ & $0.58 \pm 0.17$\\
		$B^+\to \pi^+ K^0$ &$ 23.7\pm 0.8$ & $-0.017\pm 0.016 $ & $-$ \\
		$B^+\to K^+ \pi^0$ &$ 12.9\pm 0.5$ & $0.037\pm 0.021 $ & $-$ \\
		\hline
		$B^0_d\to \pi^+ \pi^-$ & $5.12\pm 0.19 $  &$0.31 \pm 0.05$ & $-0.67 \pm 0.06$\\
		$B^0_d\to \pi^0  \pi^0$ & $1.59\pm 0.18$
			& $0.43\pm 0.24$ & $-$ \\
				$B^+\to \pi^+ \pi^0$ &$ 5.5\pm 0.4\emph{•}$ & $0.03\pm 0.04 $ & $-$ \\
	\end{tabular}
	\caption{Experimental status of $B\to\pi K,\pi\pi$ modes \cite{PDG}. In the case of the
	$B^0_d\to\pi^0\pi^0$ branching ratio ${\mathcal Br}(\pi^0\pi^0)$, we use the average of the BaBar 
	and Belle results in Refs.~\cite{Lee13,Belle17}. The branching ratios are actually CP-averaged
	quantities.}
	\label{tab:1}
\end{table}

The ``direct" CP asymmetry 
\begin{equation}
A_{\rm CP}^{\pi^0K^0}\equiv\frac{|\bar{A}_{00}|^2-|A_{00}|^2}{|\bar{A}_{00}|^2+|A_{00}|^2} = A_{\rm CP}^{\pi^0K_{\rm S}}
\end{equation}
with $\bar{A}_{00} \equiv A(\bar{B}_d^0\to \pi^0\bar{K}^0)$ and $A_{00} \equiv A(B_d^0\to\pi^0 K^0)$ is generated
directly at the amplitude level through interference between penguin and tree contributions with both CP-violating 
weak and CP-conserving strong phase differences, which are governed by
\begin{equation}\label{rc-par}
r_{\rm c}e^{i\delta_{\rm c}} \equiv (\hat T+\hat C)/P' = (0.17\pm0.06)e^{i(1.9\pm23.9)^\circ}
\end{equation}
\vspace*{-0.7truecm}
\begin{equation}\label{r-par}
re^{i\delta} \equiv (\hat T-\hat P_{tu})/P' = (0.09\pm0.03)e^{i(28.6\pm21.4)^\circ}.
\end{equation}
These parameters measure ratios of tree to penguin amplitudes, with $\hat P_{tu}$ describing the 
difference of penguin topologies with top- and up-quark exchanges. The numerical values follow from 
the strategy of Refs.~\cite{BFRS-1,BFRS-2}, utilizing the $B\to\pi\pi$ 
data in Table~\ref{tab:1} and the $SU(3)$ flavor symmetry. We include $SU(3)$-breaking corrections, assuming
non-factorizable effects of $20 \%$ \cite{FJV-B,FJV-S}, and experimental uncertainties \cite{Big}. 

Since $B^+\to\pi^+K^0$ has no color-allowed/-suppressed tree contributions which could interfere with 
penguin amplitudes, only a tiny direct CP asymmetry could originate in the SM from ${\cal O}(\lambda^2)$ terms.
Neglecting them yields
\begin{equation}\label{A+0}
A(B^+\to\pi^+K^0)= - P' = A(B^-\to\pi^-\bar K^0).
\end{equation} 
Direct CP asymmetries of other $B\to\pi K$ decays arise at ${\cal O}(r_{(c)})$.  The data in 
Table~\ref{tab:1} agree with Eq.~(\ref{A+0}) and the pattern expected from Eqs.~(\ref{rc-par}) and (\ref{r-par}). 
Keeping only terms linear in $r_{\rm (c)}$ yields the following sum rule \cite{GR,gro}:  
\begin{displaymath}
\Delta_{\rm SR} \equiv  
\left[A^{\pi^+ K^0}_{\rm CP}\frac{{\mathcal Br}(\pi^+ K^0)}{{\mathcal Br}(\pi^- K^+)}
-A^{\pi^0K^+}_{\rm CP}
\frac{2{\mathcal Br}(\pi^0 K^+)}{{\mathcal Br}(\pi^- K^+)}\right]\frac{\tau_{B_d}}{\tau_{B^\pm}}
\end{displaymath}
\vspace*{-0.7truecm}
\begin{equation}\label{Del-SR}
+A^{\pi^-K^+}_{\rm CP} - A^{\pi^0K^0}_{\rm CP}\frac{2{\mathcal Br}(\pi^0K^0)}{{\mathcal Br}(\pi^-K^+)} 
= {\cal O}(r_{(c)}^2).
\end{equation}
Using the measurements in Table~\ref{tab:1}, this relation gives
\begin{equation}\label{Adir-pred}
A^{\pi^0K^0}_{\rm CP}=-0.14\pm 0.03.
\end{equation}

\begin{figure}[tbp]    \centering
   \includegraphics[width=1.5in]{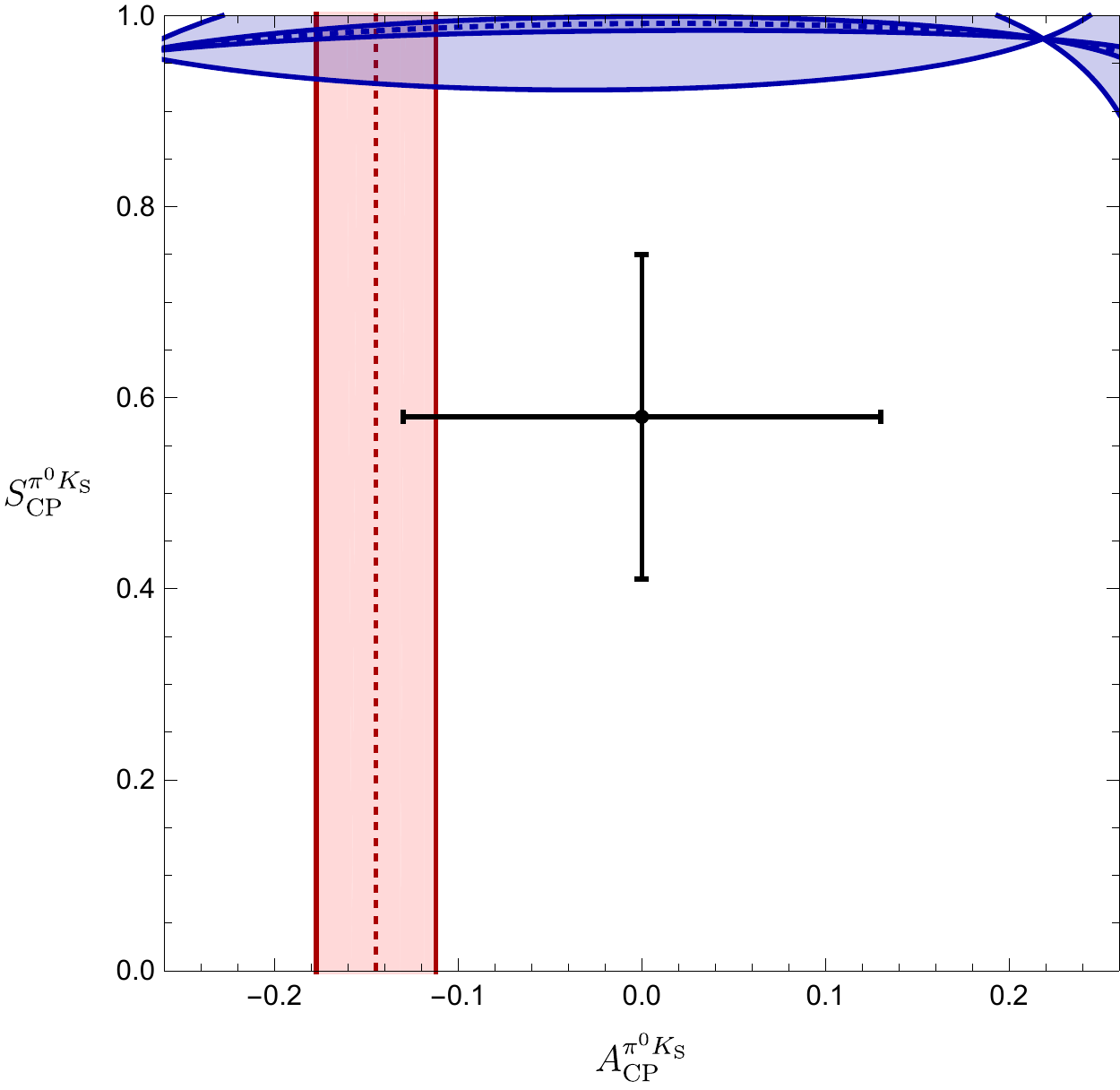} 
    \includegraphics[width=1.5in]{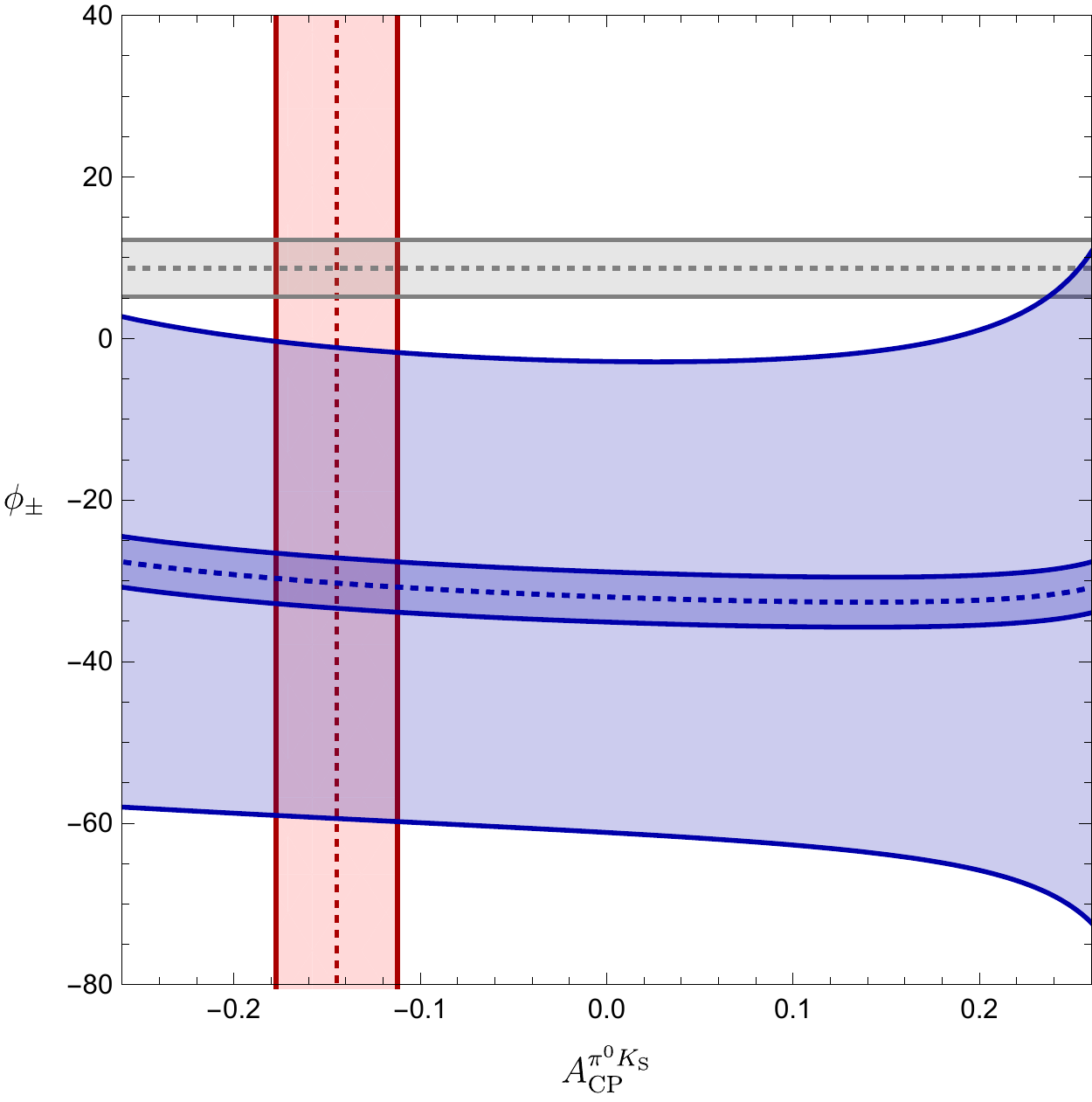} 
   \caption{Left: SM correlation between the CP asymmetries of $B^0_d\to\pi^0K_{\rm S}$, showing also 
   Eq.~(\ref{Adir-pred}) and the current data. Right: $\phi_\pm$ corresponding to the correlation in the
   left panel and comparison with the theoretical value in Eq.~(\ref{phi-pm}). The narrow bands indicate 
   the future theory uncertainties.}
   \label{fig:1}
\end{figure}

Let us now return to mixing-induced CP violation in $B^0_d\to\pi^0K_{\rm S}$. In line with Refs.~\cite{FJPZ,FFM}, 
we write
\begin{equation} \label{eq:spiK}
(\sin2\beta)_{\pi^0K_{\rm S}}\equiv \frac{S^{\pi^0K_{\rm S}}_{\rm CP}}{\sqrt{1- (A^{\pi^0K_{\rm S}}_{\rm CP})^2}}
=\sin(\phi_d - \phi_{00}),
\end{equation}
where 
\begin{equation}
 \phi_{00} \equiv \rm{arg}(\bar{A}_{00} A^*_{00})
\end{equation}
is the angle between $\bar{A}_{00}$ and $A_{00}$, and $\phi_d=(43.2\pm1.8)^\circ$ denotes the CP-violating 
$B^0_d$--$\bar B^0_d$ mixing phase; the numerical value follows from CP violation in 
$B^0_d\to J/\psi K_{\rm S}$ \cite{PDG}, including corrections from penguin effects \cite{DeBrFl}. 

As was pointed out in Ref.~\cite{FJPZ} (see also Ref.~\cite{groro}), a correlation between the CP 
asymmetries of $B^0_d\to \pi^0K_{\rm S}$ can be obtained with the help of the following isospin relation 
\cite{NQ,GHLR}:
\begin{displaymath}
 \hspace*{-1.5truecm} \sqrt{2} A(B^0_d \to \pi^0 K^0) + A(B^0_d \to \pi^- K^+) 
 \end{displaymath}
 \vspace*{-0.8truecm}
\begin{equation}\label{iso-rel-neut}
= - (\hat{T} +\hat{C})e^{i\gamma} + (\hat P_{EW} + \hat P_{EW}^{\rm C}) \equiv 3A_{3/2}.
\end{equation}
Here $\gamma$ is the usual angle of the UT and
\begin{equation}\label{A32}
3 A_{3/2} \equiv 3 |A_{3/2}|e^{i\phi_{3/2}} = - (\hat{T} +\hat{C})\left(e^{i\gamma} - q e^{i\phi} e^{i\omega}\right) 
\end{equation} 
denotes an isospin $I=3/2$ amplitude. Its normalization can be determined through the branching ratio 
${\mathcal Br}(\pi^+ \pi^0)$ utilizing the following $SU(3)$ relation \cite{GRL,BF-98}:
\begin{equation}\label{T+C-det}
|\hat{T}+\hat{C}|= R_{T+C}\left|\frac{V_{us}}{V_{ud}}\right|\sqrt{2} |A(B^+\to\pi^+ \pi^0)|,
\end{equation}
where $R_{T+C}\approx f_K/f_\pi =  1.2 \pm 0.2$ describes $SU(3)$-breaking effects 
\cite{BBNS,BeNe}. For given values of $A^{\pi^0K_{\rm S}}_{\rm CP}$ and $(q,\phi)$, the measured 
neutral $B\to\pi K$ branching ratios then allow us to determine $\phi_{00}$ from the amplitude triangles representing
Eq.~(\ref{iso-rel-neut}) and its CP conjugate. Finally, using Eq.~(\ref{eq:spiK}), we may calculate 
$S^{\pi^0K_{\rm S}}_{\rm CP}$.

Employing the SM values in Eq.~\eqref{q-SM} and the data in Table~\ref{tab:1}, we obtain the 
correlation in the left panel of Fig.~\ref{fig:1}. In comparison with Ref.~\cite{FJPZ}, the picture is now 
considerably more constrained, which is in particular due to an improved value of $\gamma = (70\pm 7)^\circ$. 
To further explore the underlying structure, we consider the angle
\begin{equation}\label{eq:pmcon}
\phi_\pm =  \rm{arg}(\bar{A}_\pm A_\pm^*)
\end{equation}
between the amplitudes $\bar A_\pm \equiv A(\bar B^0\to \pi^+ K^-)$ and $A_\pm \equiv A(B^0\to \pi^- K^+)$, where 
EW penguins enter only in colour-suppressed form. In the case of $\phi=0^\circ$, we have 
\begin{equation}\label{phi-pm}
\left.\phi_\pm \right |_{\phi=0}= 2 \, r \cos\delta\sin\gamma + {\cal O}(r^2)= (8.7 \pm 3.5)^\circ,
\end{equation}
where we used Eq.~(\ref{r-par}). In the right panel of Fig.~\ref{fig:1}, we show the range for $\phi_\pm$ following from 
the contour in the left panel. The present uncertainty is driven by the precision of $\mathcal Br(\pi^0K^0)$ and 
$R_q$. We add a future scenario with perfect measurements, assuming that progress in theory will give 
$R_q = 1.00\pm 0.05$ \cite{FJPZ}. We show also Eq.~(\ref{phi-pm}) as the horizontal 
band. Surprisingly, it disagrees with the experimental constraint, which is a new aspect of the $B\to\pi K$ puzzle. 
Consequently, both the data for CP violation in $B^0_d\to\pi^0K_{\rm S}$ and the correlation itself 
are in conflict with the SM. This puzzling picture would then require data to move, with in particular 
${\mathcal Br}(\pi^0K^0)$ going down by about $2.5\,\sigma$ and changes of the CP asymmetries at the 
$1\sigma$ level \cite{Big}. However, the tension in the data may also indicate a modified EW penguin sector. 

This intriguing situation is not yet conclusive. Fortunately, the Belle II experiment offers exciting prospects for much 
more precise measurements of $B^0_d\to \pi^0K^0$ \cite{Belle-II}. Moreover, thanks also to  LHCb \cite{LHCb}, 
further parameters will be known with much higher precision, such as $\gamma$ with uncertainties at the $1^\circ$ 
level. Belle II has already performed a feasibility study for the sum rule in Eq.~(\ref{Del-SR}), where the key limitation 
is due to the experimental uncertainty of $\Delta A^{\pi^0K^0}_{\rm CP}|_{\rm exp} \sim 0.04$ \cite{Belle-II}. 

Could this analysis resolve NP contributions to the EW penguin sector? Using the exact expressions in 
Ref.~\cite{BFRS-2} with the parameters in Eqs.~(\ref{rc-par}--\ref{r-par}), we obtain
\begin{equation}\label{DelSR-expr}
\left.\Delta_{\rm SR}\right|_{\rm SM} = -0.009 \pm 0.014
\end{equation}
for the SM values in Eq.~(\ref{q-SM}). 
In order to detect a deviation from Eq.~(\ref{DelSR-expr}) at the $1\sigma$ level at Belle II, we find that the 
NP effects must at least be as large as $q\sim3$ with $\phi\sim160^\circ$. In view of the global pattern of the 
current data, we focus on $q<3$, including the SM regime. For much larger values of  $q$, the significance could 
grow, depending on the specific value of $\phi$. However, even then additional information would be necessary 
to resolve the underlying NP effects.

In view of this situation, we have developed a new strategy to determine $q$ and $\phi$ from $B \to \pi K$ decays. 
The starting point is given by the isospin relation in Eq.~(\ref{iso-rel-neut}) and its counterpart for the charged 
$B\to\pi K$ decays:
\begin{equation}\label{iso-rel-char}
\sqrt{2} A(B^+\to \pi^0 K^+) + A(B^+\to \pi^+ K^0) = 3A_{3/2}.
\end{equation}
Considering a given value of $|A_{3/2}|$, we may apply these relations to determine 
\begin{equation}
\Delta\phi_{3/2}\equiv \phi_{3/2}-\bar\phi_{3/2}
\end{equation} 
separately for the neutral and charged $B\to\pi K$ decays from the corresponding amplitude triangles. 
Employing Eq.~(\ref{T+C-det}), we convert $|A_{3/2}|$ into
\begin{equation}\label{N-def}
N\equiv 3 |A_{3/2}|/|\hat{T} +\hat{C}|,
\end{equation}
where we used the model-independent result of a small $\omega$, which implies $|A_{3/2}|=|\bar A_{3/2}|$. The 
numerical analysis turns out to be essentially insensitive to $\omega$ for values as large as about $10^\circ$. The 
following expressions allow us then to calculate contours in the $\phi$--$q$ plane:
\begin{equation}\label{q-det}
q=\sqrt{N^2-2c\cos\gamma-2s\sin\gamma+1},
\end{equation}
\vspace*{-0.3truecm}
\begin{equation}
\tan\phi=\frac{\sin\gamma-s}{\cos\gamma-c}, \quad q\,\sin\phi=\sin\gamma-s,
\end{equation}
where 
\begin{equation}
c\equiv \pm N\cos(\Delta\phi_{3/2}/2), \quad s\equiv \pm N\sin(\Delta\phi_{3/2}/2).
\end{equation}

The practical implementation requires us to fix the relative orientation of the amplitude triangles of the decays and their 
CP conjugates. In the case of the neutral modes, this can be accomplished in a clean way through the measured value 
of $S^{\pi^0K_{\rm S}}_{\rm CP}$. For the charged decays, we  use 
Eq.~(\ref{A+0}). Employing $SU(3)$ methods, even the tiny angle between the $B^+\to \pi^+K^0$ and 
$B^-\to \pi^-\bar K^0$ amplitudes arising from ${\cal O}(\lambda^2)$ corrections can be taken into 
account \cite{FJV-B,FJV-S}, as done in our numerical analysis.

\begin{figure}[tbp] 
   \centering
     \includegraphics[width=1.6in]{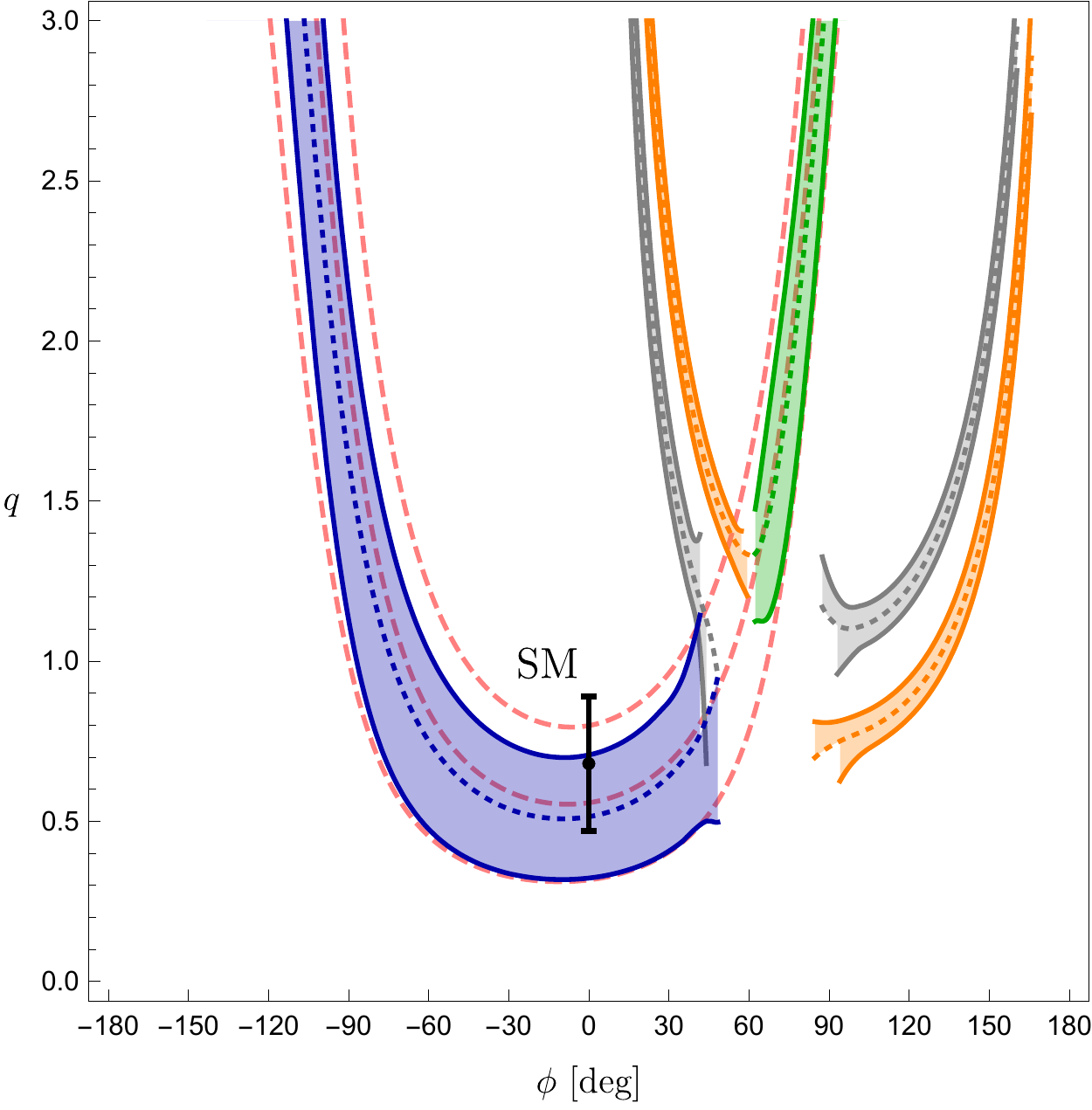} 
      \includegraphics[width=1.6in]{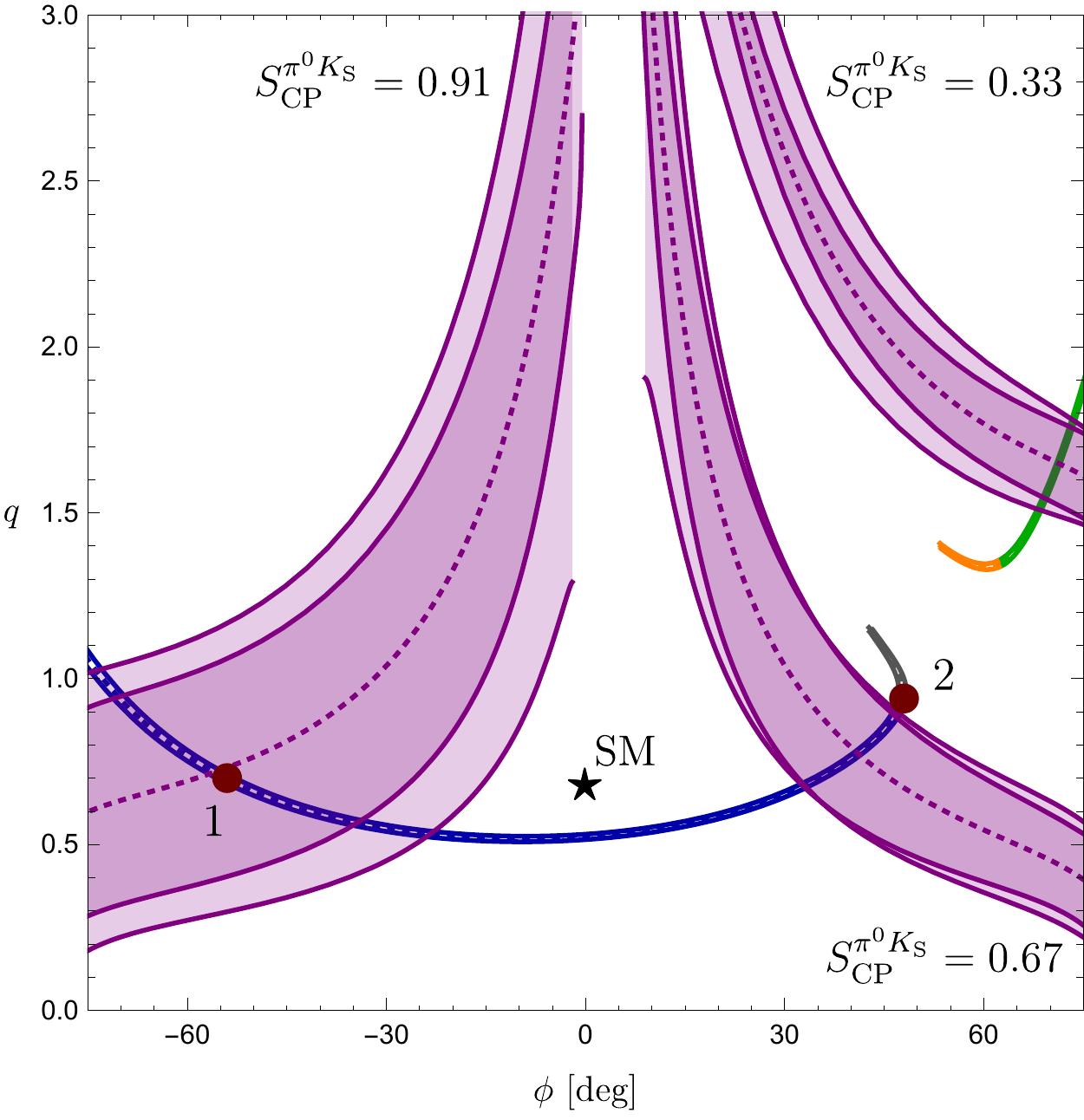} 
   \caption{Left: contours in the $\phi$--$q$ plane for charged $B\to\pi K$ data, showing also the $R_{\rm c}$
   constraint (dashed) and the SM prediction. Right: three future scenarios utilizing mixing-induced CP violation in 
   $B^0_d\to\pi^0K_{\rm S}$, as discussed in the text.}
   \label{fig:2}
\end{figure}

In view of the large current uncertainties affecting the experimental value of $S^{\pi^0K_{\rm S}}_{\rm CP}$, we 
apply the new method only to the charged $B\to\pi K$ decays. In Fig.~\ref{fig:2}, we show the contours 
satisfying all constraints, including the angle between the $B^\pm\to\pi^0K^\pm$ amplitudes. 
As the triangles require $|A_{3/2}|$ to be bigger than the lower bound following from Eq.~\eqref{iso-rel-char}, 
we encounter the discontinuities around $q\sim 1$ and $\phi\sim 70^\circ$. This analysis is very robust from the
theoretical point of view \cite{FJPZ}, relying on an isospin relation with minimal $SU(3)$ input given 
by $R_{T+C}$ in Eq.~(\ref{T+C-det}). In particular, no decay topologies, such as color-suppressed EW penguins, 
have to be neglected. 

It is useful to complement this analysis with
\begin{equation}\label{Rc-def}
R_{\rm c}\equiv 2 \left[\frac{{\mathcal Br}(\pi^0 K^+)}{{\mathcal Br}(\pi^+K^0)}\right]= 1.09\pm0.06,
\end{equation}
where the general expressions in Ref.~\cite{BFRS-2} yield
\begin{equation}\label{Rc-expr}
R_{\rm c}=  1- 2 \, r_{\rm c}\cos\delta_{\rm c}(\cos\gamma-q\cos\phi)+{\cal O}(r_{\rm c}^2).
\end{equation}
Using the parameters in Eq.~(\ref{rc-par}), $R_{\rm c}$ can be converted into another curve in the $\phi$--$q$ plane. 
As it requires information on $\delta_{\rm c}$ and the ${\cal O}(r_{\rm c}^2)$ terms involve color-suppressed 
EW penguins, it is not as clean as the triangle contours. However, as can be seen in Fig.~\ref{fig:2}, we find remarkable agreement with one of them, 
allowing for a potentially large CP-violating phase $\phi$ while being also consistent with $\phi=0^\circ$ as in the SM.

In order to reveal the full picture, further information is needed. It is offered by the mixing-induced CP asymmetry of 
$B^0_d\to\pi^0K_{\rm S}$, which allows the determination of $\phi_{00}$ from Eq.~(\ref{eq:spiK}). This angle is
given as 
\begin{displaymath}
\hspace*{-3.0truecm}\tan\phi_{00} =  2\,(r\cos\delta-r_{\rm c}\cos\delta_{\rm c})\sin\gamma
\end{displaymath}
\vspace*{-0.8truecm}
\begin{equation}\label{tanphi00}
+2\, r_{\rm c} \left(\cos\delta_{\rm c}-2 \, \tilde a_{\rm C} /3 \right) q\sin\phi +{\cal O}(r_{\rm (c)}^2),
\end{equation}
where $\tilde a_{\rm C} \equiv a_{\rm C}\cos(\Delta_{\rm C} + \delta_{\rm c})$
with the CP-conserving strong phase $\Delta_{\rm C}$ describes color-suppressed 
EW penguins \cite{BFRS-2}. Since these contributions, having a ``naive" value of $a_{\rm C}\sim0.2$, may
be enhanced by non-perturbative effects, we employ data to take them into account, using the following 
ratio \cite{FM}:
\begin{equation}\label{R-def}
\hspace*{-0.2truecm}R \equiv \left[\frac{{\mathcal Br}(\pi^- K^+)}{{\mathcal Br}(\pi^+K^0)}\right]
\frac{\tau_{B^\pm}}{\tau_{B_d}} = 0.89 \pm 0.04,
\end{equation}
which can be written as
\begin{equation}
R=1-2\,r\cos\delta\cos\gamma+2 \, r_{\rm c} \, \tilde a_{\rm C} \, q \cos\phi +{\cal O}(r_{\rm (c)}^2),
\end{equation}
where we have assumed Eq.~(\ref{A+0}). We observe that $R$ allows us to include the color-suppressed 
EW penguin contribution to Eq.~(\ref{tanphi00}). As we noted above, we may take corrections to Eq.~(\ref{A+0})
into account, as we do in the numerical analysis. Using Eqs.~(\ref{rc-par}) and (\ref{r-par}) yields then
\begin{equation}
\tilde a_{\rm C} \, q \cos\phi =  -0.10 \pm 0.15,\quad \tilde a_{\rm C}|_{\rm SM}= -0.15 \pm  0.23,
\end{equation}
indicating a small impact of the color-suppressed EW penguin topologies in $R$ and $\phi_{00}$.
Higher order terms in $r_{\rm (c)}$ can also be included, involving terms of the form 
$a_{\rm C}\sin(\Delta_{\rm C}+\delta_c)$, which can be constrained through the direct CP asymmetry of 
$B^0_d\to\pi^-K^+$.

\begin{figure}[t]
   \centering
    \includegraphics[width=1.6in]{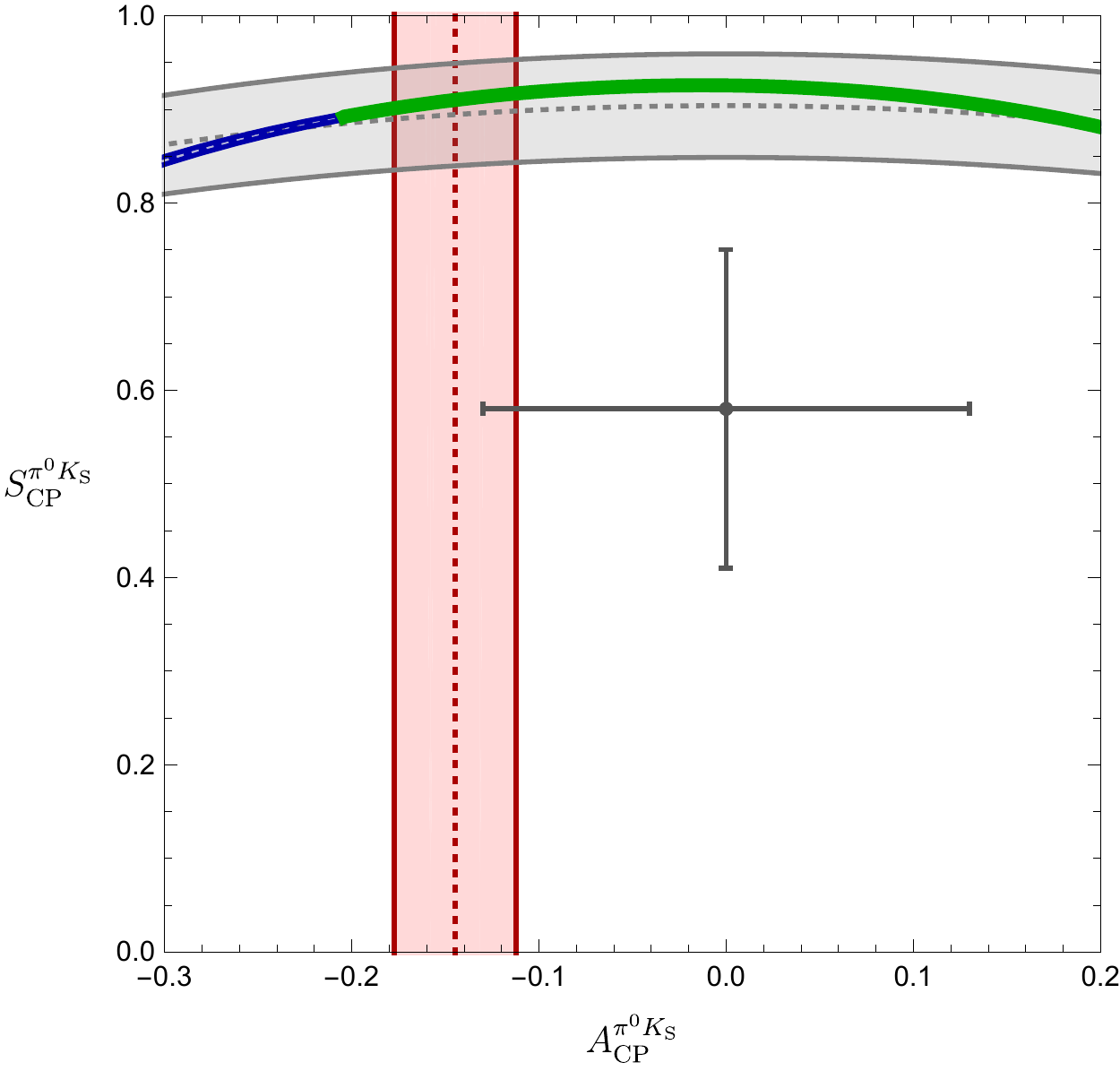} 
    \includegraphics[width=1.6in]{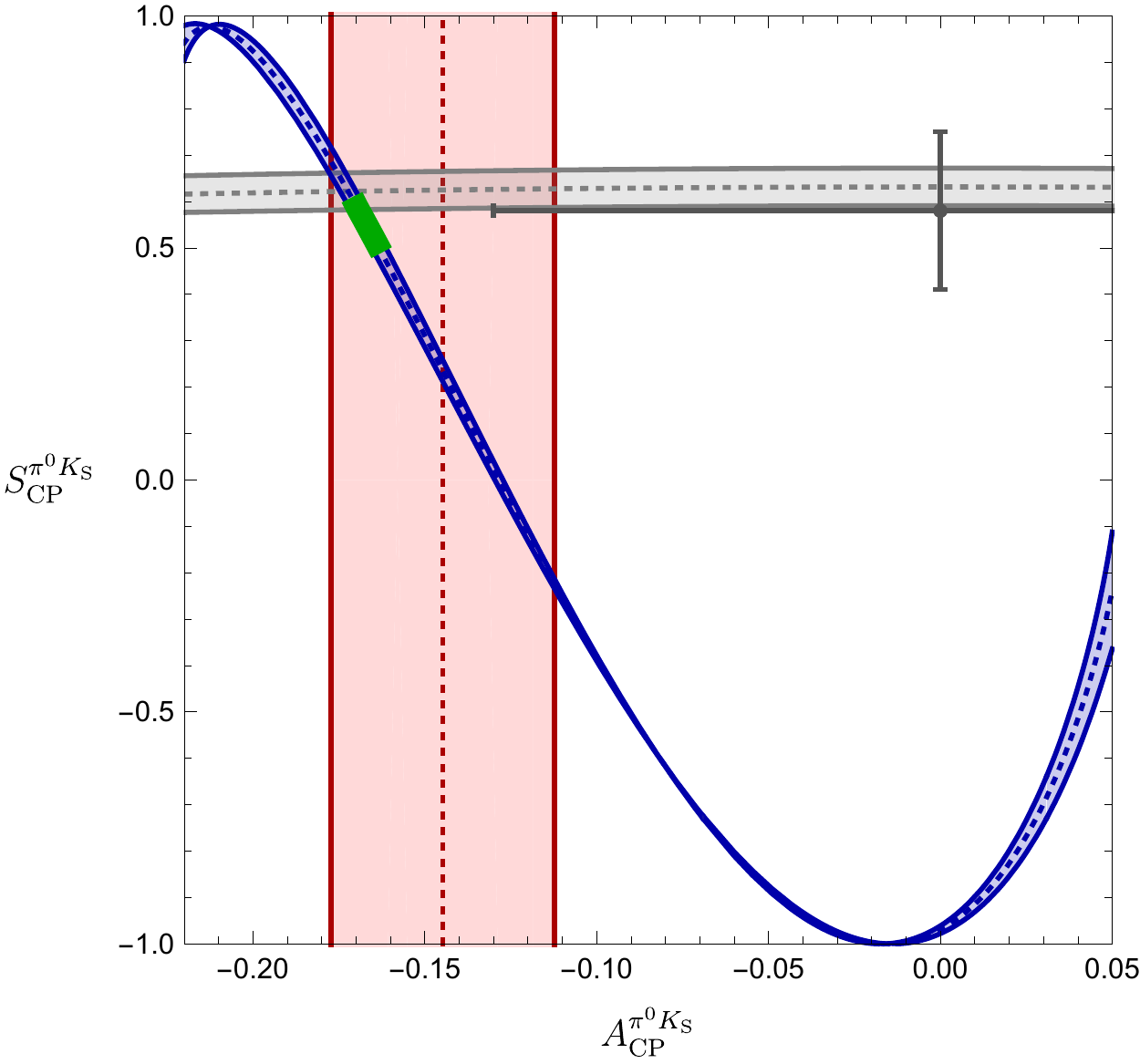} 
   \caption{Correlations between $A^{\pi^0K_{\rm S}}_{\rm CP}$ and $S^{\pi^0K_{\rm S}}_{\rm CP}$ for the NP 
   scenarios labelled by ``1" (left) and ``2" (right) in Fig.~\ref{fig:2}. For the green parts of the contours, the value 
   of $\phi_\pm$ agrees with Eq.~(\ref{phi-pm}). The grey horizontal bands correspond to Eq.~(\ref{eq:spiK}),
   calculating  $\phi_{00}$ with the parameters in Eqs.~(\ref{rc-par}) and (\ref{r-par}).}\label{fig:3}
\end{figure}

In the case of $\phi=0^\circ$, the EW penguin contribution in Eq.~(\ref{tanphi00}) vanishes while these topologies 
could still enter through tiny $r_{\rm c}^2$ terms. Consequently, $\phi_{00}$ is then essentially insensitive to EW 
penguin effects, and we obtain the prediction $(\sin2\beta)_{\pi^0K_{\rm S}} = 0.80 \pm 0.06$, where the error takes 
$SU(3)$-breaking effects of $20\%$ and ${\cal O}(r_{\rm c}^2)$ corrections into account.

Let us now allow for large CP-violating phases $\phi$. In the right panel of Fig.~\ref{fig:2}, we discuss a future 
scenario, assuming theory uncertainties as in Ref.~\cite{FJPZ}. We show only triangle contours which are consistent 
with the $R_{\rm c}$ constraint. We consider three sets of $(q,\phi)$. In the corresponding $\phi_{00}$ curves,  
we take color-suppressed EW penguin contributions through $R$ and $A^{\pi^-K^+}_{\rm CP}$ into account, and 
show both the theoretical (wide bands) and experimental (small bands) uncertainties. The former correspond to 
non-factorizable $SU(3)$-breaking effects of $20\%$, although we expect a much sharper picture of 
$SU(3)$-breaking effects by the time these measurements will be available through data-driven methods and 
progress in theory \cite{FJV-B,FJV-S}. Since Belle II did not give uncertainties for $S^{\pi^0K_{\rm S}}_{\rm CP}$, 
we assume a resolution as for $\Delta A^{\pi^0K^0}_{\rm CP}$ \cite{Belle-II}. We observe that the theoretical 
uncertainties of the $\phi_{00}$ contours would then allow us to match the corresponding experimental precision, which is very promising.

Finally, we return to the $B\to\pi K$ puzzle. Is it actually possible to resolve it through values of $q$ and $\phi$ 
in agreement with the allowed regions following from the charged $B\to\pi K$ data? In Fig.~\ref{fig:3}, we zoom into 
the $A^{\pi^0K_{\rm S}}_{\rm CP}$--$S^{\pi^0K_{\rm S}}_{\rm CP}$ plane for two NP scenarios, showing only 
triangle contours which satisfy the internal consistency requirements. Evidently, the $B\to\pi K$ puzzle could be 
resolved, which is very non-trivial in view of the  subtle features involved and the strong constraints for $q$ and 
$\phi$ implied by the charged $B\to\pi K$ data. Further insights can be obtained through the 
independent triangle analysis for the neutral $B\to\pi K$ decays once the experimental uncertainty for 
$S^{\pi^0K_{\rm S}}_{\rm CP}$ can be reduced. This observable will play the key role in the future exploration of the 
$B\to\pi K$ system.

It will be very interesting to see whether the implementation of our new strategy at Belle II will once again confirm the 
SM or finally establish NP in the EW penguin sector with possible new sources of CP violation. In the latter case, 
the corresponding new particles, such as additional $Z'$ bosons, may be linked to anomalies in current data for rare 
semileptonic $B$ decays, thereby offering an exciting new playground for future explorations.

%
%
%
\section*{Acknowledgements} 
We would like to thank Eleftheria Malami for discussions. This research has been supported by the Netherlands 
Organisation for Scientific Research (NWO) and by the Deutsche Forschungsgemeinschaft (DFG) within research 
unit FOR 1873 (QFET). K.K.V. acknowledges hospitality and support for her regular visits of Nikhef. 
%
%
%

%
%
%
\end{document}